\documentclass[prl,superscriptaddress,onecolumnpage,notitlepage]{revtex4-1}

\pdfoutput=1
\usepackage{graphicx} 
\usepackage{tikz}
\usepackage{amsmath}
\usepackage{amssymb}
\usepackage{natbib}
\usepackage{color}
\usepackage{float}
\usepackage{epstopdf}
\usepackage[normalem]{ulem}
\usepackage{pgfplots,pgfplotstable}
\usepackage{soul}
\usepackage{accents}
\usepackage{nameref}
\usepackage{balance}
\usepackage[all]{nowidow}

\newcommand{\id}{\mathrm{d}} 
\newcommand{\vett}[1]{\mathbf{#1}} 

\newcommand{\subfigimg}[3][,]{%
	\setbox1=\hbox{\includegraphics[#1]{#3}}
	\leavevmode\rlap{\usebox1}
	\rlap{\hspace*{5pt}\raisebox{\dimexpr\ht1-0\baselineskip}{#2}}
	\phantom{\usebox1}
}

\begin{document}
\title{Snapping of Bistable, Prestressed Cylindrical Shells}
\author{Xin Jiang}
\affiliation{
Department of Mechanical Engineering, Boston University, Boston, MA, 02215.
}%

\author{Matteo Pezzulla}
\affiliation{
Department of Mechanical Engineering, Boston University, Boston, MA, 02215.
}%

\author{Huiqi Shao}
\affiliation{
Textile Engineering, Chemistry, and Science Department, North Carolina State University, Raleigh, NC, 27695.
}%

\author{Tushar K. Ghosh}
\affiliation{
	Textile Engineering, Chemistry, and Science Department, North Carolina State University, Raleigh, NC, 27695.
}%

\author{Douglas P. Holmes}
\email{dpholmes@bu.edu}
\affiliation{
Department of Mechanical Engineering, Boston University, Boston, MA, 02215.
}%

\date{\today}

\begin{abstract}
Bistable shells can reversibly change between two stable configurations with very little energetic input. Understanding what governs the shape and snap--through criteria of these structures is crucial for designing devices that utilize instability for functionality. Bistable cylindrical shells fabricated by stretching and bonding multiple layers of elastic plates will contain residual stress that will impact the shell's shape and the magnitude of stimulus necessary to induce snapping. Using the framework of non--Euclidean shell theory, we first predict the mean curvature of a nearly cylindrical shell formed by arbitrarily prestretching one layer of a bilayer plate with respect to another. Then, beginning with a residually stressed cylinder, we determine the amount of the stimuli needed to trigger the snapping between two configurations through a combination of numerical simulations and theory. We demonstrate the role of prestress on the snap--through criteria, and highlight the important role that the Gaussian curvature in the boundary layer of the shell plays in dictating shell stability.
\end{abstract}

\pgfplotsset{compat=newest}

\maketitle

Multistable structures made with soft materials can reversibly change between stable configurations through a snap--through elastic instability. Snap--through is a limit point instability~\cite{thompson1984elastic} that is commonly observed in the eversion of an umbrella on a windy day, or in the jumping of a toy popper~\cite{pandey2013swelling}. Such structures have utility in engineering and material systems due to their ability to remain in an alternate configuration following the removal of the applied stimulus. Bistable structures have been used in the design of biomedical devices~\cite{goncalves2003}, micro--electromechanical systems \cite{hsu2003,das2009}, energy harvesters~\cite{pellegrini2013bistable,harne2013review}, morphing structures \cite{thill2008,fernandes2010,pirrera2010}, and architected materials that trap strain energy~\cite{shan2015multistable}. The snap--through of these bistable systems can be triggered by a wide range of stimuli, including temperature~\cite{jakomin2010}, light~\cite{shankar2013contactless} and swelling~\cite{xia2010,holmes07,pezzulla2017global}.

Recent research has focused on the snap--through of shells induced by non--mechanical stimuli, such as an evolving natural curvature~\cite{pezzulla2017curvature,pezzulla2017global}. Such structures often do not possess a stress--free equilibrium configuration, and are therefore modeled using a non--Euclidean shell theory~\cite{efrati2009}. The critical natural curvature needed to induce snapping of a bistable, stress--free cylindrical shell was recently shown to be proportional to the shell's initial curvature~\cite{pezzulla2017curvature}. The snap--through of prestressed cylindrical shells in response to a mechanical force is commonly encountered in structures such as tape springs, tape measures, and toy snap--bracelets~\cite{kebadze2004bistable}, but little is understood about the response of these shells to non--mechanical stimuli. Non--mechanical loading of prestressed shells is particularly relevant to electrically active polymers (EAP) wherein dielectric elastomers are deformed in response to an applied voltage~\cite{carpi2011bioinspired}.

The voltage--induced stretching of EAPs is inversely proportional to the material's thickness~\cite{pelrine2000high}, which has led researchers to significantly prestretch the nearly incompressible dielectric elastomers to reduce their thickness. When EAPs are made of multiple active and passive layers, applying voltage to one active layer will differentially stretch that layer with respect to the rest of the material, causing it to bend with a voltage--induced natural curvature. This process is analogous to the heating of bimetallic strips wherein temperature generates a natural curvature in the beam. Maintaining either a large stretch or bending deformation in a dielectric elastomer requires the continued application of an applied voltage, which has a high energetic cost and often leads to material failure through dielectric breakdown~\cite{huang2012thickness,trols2013stretch}. While this material instability is irreversible, reversible elastic instabilities can also be triggered, including creasing and wrinkling instabilities in the presence of surface tension~\cite{park2013,zhao2007electromechanical}, bending~\cite{christianson2017fluid} and buckling~\cite{tavakol2016} using fluid electrodes, and snap--through instabilities in the presence of pressure~\cite{zhao2014harnessing,zhao2010,suo2008,suo2010}. Elastic instabilities present an alternate means for generating functionality in EAPs, as they enable soft, multistable structures to reversibly snap between configurations triggered by voltage only applied for a short amount of time. 

In this Letter, we examine the bistability of prestressed shells loaded by a non--mechanical stimulus. We will first generalize the stimulus as one that induces a stretch in the material, and describe how the magnitude of prestretch applied to a flat, bilayer plate dictates the curvature of the resulting prestressed shell. Second, we will identify the geometric criteria for inducing snap--through of a prestressed shell. To accomplish the first step, we establish relationships between the prestretch and the natural curvature, and then determine the relationship between the natural curvature and the resulting shell's mean curvature using non--Euclidean shell theory. Then, we will examine how much stretch--inducing stimulus, {\em e.g.} voltage, is needed to induce snapping of these shells as a function of geometry and prestress. 

\begin{figure} \vspace{-4mm}
	\begin{center}
		\begin{tabular}{@{}p{0.75\columnwidth}@{\quad}p{width=1\columnwidth}@{}}
			\subfigimg[width=0.75\columnwidth]{$a.$}{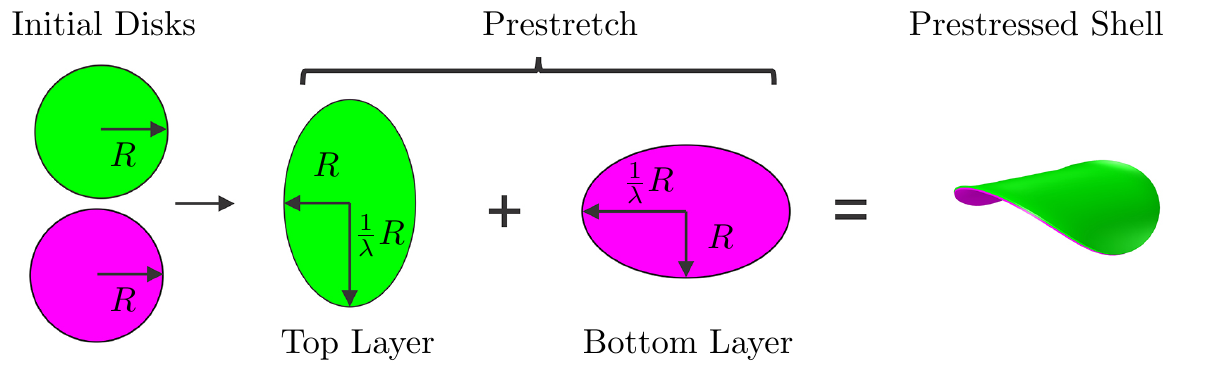} 
			\subfigimg[width=0.75\columnwidth]{$b.$}{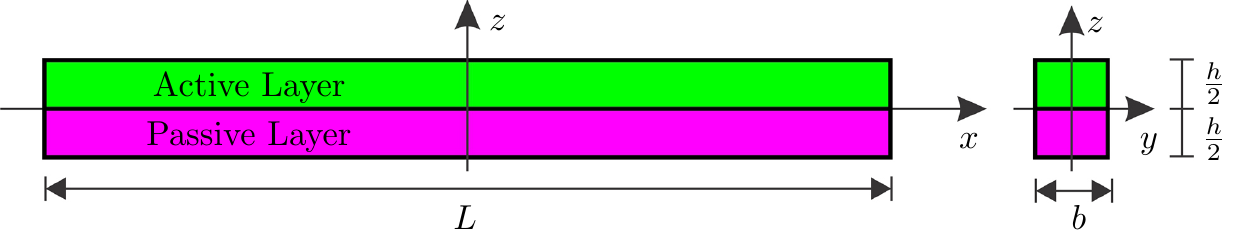}		
			\subfigimg[width=0.67\columnwidth]{$c.$}{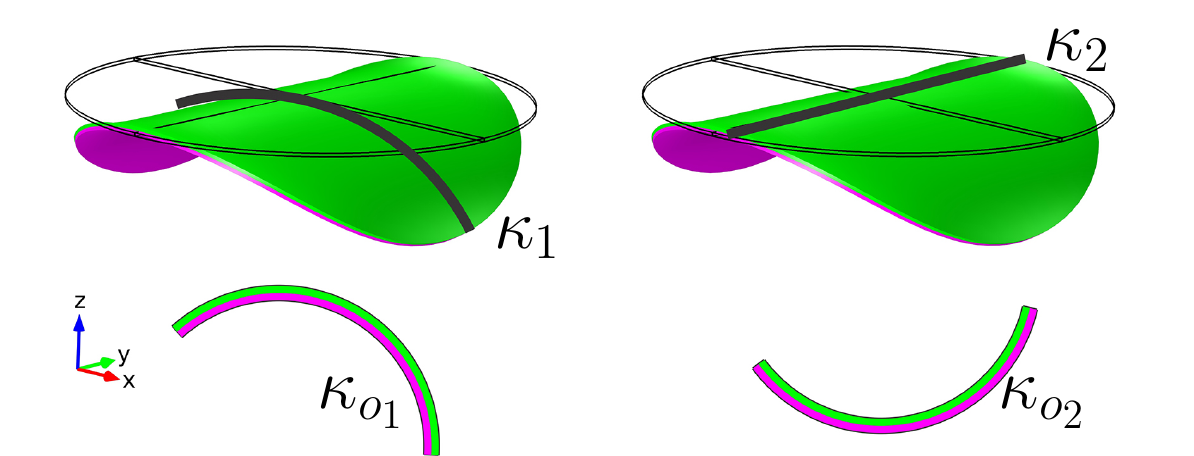}
		\end{tabular}				
	\end{center}
	\vspace{-5mm}	
	\caption{$a.$ The fabrication of the bistable shell. In practice, two rectangular sheets were first stretched and bonded together, then a circular disk is cut from the bilayer sheets. $b.$ Illustration of a bilayer beam. The top layer is active while the bottom layer is passive. $c.$ Two beams cut from orthogonal directions of the shell. $\kappa_{o_1}$ and $\kappa_{o_2}$ are the stress free natural curvatures of the two beams.}
	\label{fig1}\vspace{-5mm}
\end{figure}

Consider two dielectric elastomer plates that are stretched in orthogonal directions and bonded together (Fig.~\ref{fig1}). Upon release of the prestretch, the disks will buckle into a bistable, prestressed cylindrical shell with a mean curvature $\mathcal{H}=1/2(\kappa_1+\kappa_2)$~\footnote{Experimentally, this can be accomplished by stretching two rectangular disks, bonding them together, then cutting a circular shape from the bilayer disk before releasing the external force.}. Since the stretching energy of a thin shell is proportional to the thickness $h$, while the bending energy is proportional to $h^3$, it is generally suitable to determine the mean curvature of the deformed shell by assuming it deforms isometrically, and minimizing the shell's bending energy. This approach was previously used to determine the deformed shape of bilayer plates where one layer was subjected to a homogenous, spherical curvature--inducing stimulus~\cite{pezzulla2016geometry,pezzulla2017curvature}. This work found that in the isometric limit, the mean curvature of the resulting cylinder is given by
\begin{equation}
	\label{meanSpherical}
	\mathcal{H}=\frac{1+\nu}{2}\kappa_\textup{o},
\end{equation}
where $\nu$ is Poisson's ratio, and $\kappa_\textup{o}$ is the natural curvature. The natural curvature can be thought of as the curvature a 1D beam would adopt if cut from the cylindrical shell's surface. The 1D object can exactly adopt the curvature stimulus, while the plate or shell in general cannot due to conditions required for geometric compatibility. This problem was later generalized to a non--spherical curvature--inducing stimulus, with a curvature--inducing stimulus being applied in two orthogonal, principal directions, and the mean curvature of the resulting cylinder was found to be~\cite{pezzulla2017curvature}
\begin{equation} 
	\label{H_kappa_a_b}
	\mathcal{H}=\frac{1}{2} (\kappa_{o_1}+\nu \kappa_{o_2}) \,,
\end{equation}
where $\kappa_{o_1}$ and $\kappa_{o_2}$ are the natural curvature of the two beams cut from the shell in principal directions (Fig. \ref{fig1}). We note that equation~\ref{H_kappa_a_b} reduces to equation~\ref{meanSpherical} when curvature--inducing stimulus is spherical, {\em i.e.} $\kappa_{o_1}=\kappa_{o_2}=\kappa_\textup{o}$. 

Describing the mean curvature of a residually stressed shell in terms of the magnitude of the applied, curvature inducing stimuli is useful if the magnitude of this stimulus is known. Previous studies have used the residual swelling of elastomeric plates as a means for inducing curvature in an object~\cite{Pezzulla2015}, and the magnitude of that residual curvature was determined by measuring the curvature of a bilayer beam of the same material and thickness as the plates~\cite{pezzulla2016geometry, pezzulla2017instabilities}. In general, this approach may not be possible. Here, we seek a relationship between the stretch applied to the plates and the resulting cylindrical shell's mean curvature. The relationship between the swelling--induced stretch of one layer relative to another was determined by Lucantonio {\em et al.} in the context of a bilayer beam with one active layer that swells by $\lambda$, and one passive layer~\cite{lucantonio2014swelling}. For simplicity, we adapt the expression for $\kappa_\textup{o}$ as a function of $\lambda$ that was given for bilayers with arbitrary thickness and moduli ratios~\cite{lucantonio2014swelling, pezzulla2016geometry} to a beam with layers of equal thickness and equal modulus, which leads to
\begin{equation}
	\label{kappao}
	\kappa_\textup{o}=-\frac{3}{8h} \frac{1+\lambda(\lambda+14)}{(\lambda+1)^2} \frac{\lambda-1}{\lambda} \,,
\end{equation}
where $\kappa_\textup{o}$ is the natural curvature of a bilayer beam under homogeneous stretch. As expected, the final curvature of the beam is proportional to~$1/h$. 

The prestrain of nearly incompressible dielectric elastomers in the formation of electrically active polymers can be over 100\%, leading to a non--negligible thickness change due to the Poisson effect. This thickness change will affect the natural curvature, and in the case of homogeneous stretch of the active layer, the final thickness will be $h_\textup{t}=\frac{h}{2}(\lambda+1)$, while the final thickness of the passive layer will be $h_\textup{b}=\frac{h}{2}(\lambda^{-1}+1)$ due to the Poisson effect. Therefore when the active layer is incompressible, the natural curvature of the beam is:
\begin{equation} \label{kappa_o12}
	\kappa_{o_1}=-\kappa_{o_2}=\mathcal{F}(\lambda)=\frac{h_\textup{t}}{h_\textup{b}} \kappa_\textup{o}=\frac{1+\lambda}{1+ \lambda^{-1}} \kappa_\textup{o}\,.
\end{equation} 
From equations~(\ref{H_kappa_a_b}) and (\ref{kappa_o12}), we find that the mean curvature of a prestressed cylindrical shell as a function of $\lambda$ and accounting for a change in thickness will be
\begin{equation} 
	\label{H_final}
	H=\frac{1-\nu}{2} \left(\frac{1+\lambda}{1+ \lambda^{-1}}\right)\kappa_\textup{o}\,.
\end{equation}

\begin{figure} \vspace{-4mm}
	\begin{center}
		\begin{tabular}{@{}p{0.76\columnwidth}@{\quad}p{width=1\columnwidth}@{}}
			\subfigimg[width=0.70\columnwidth]{$a.$}{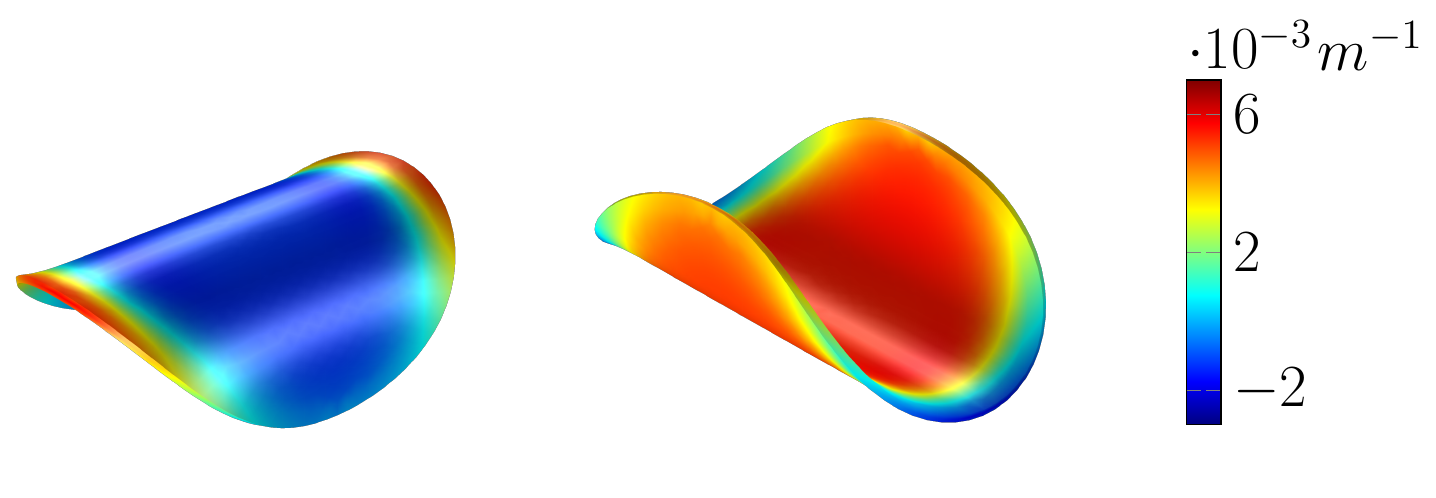}					
			\subfigimg[width=0.36\columnwidth]{$b.$}{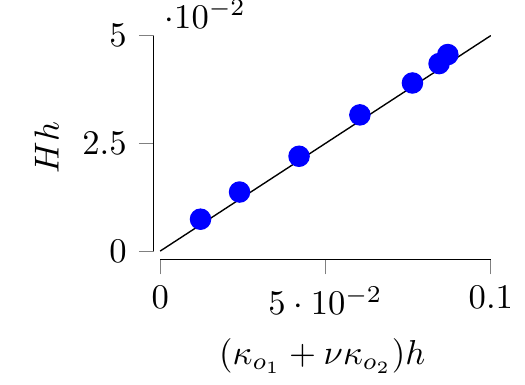}		
			\subfigimg[width=0.39\columnwidth]{$c.$}{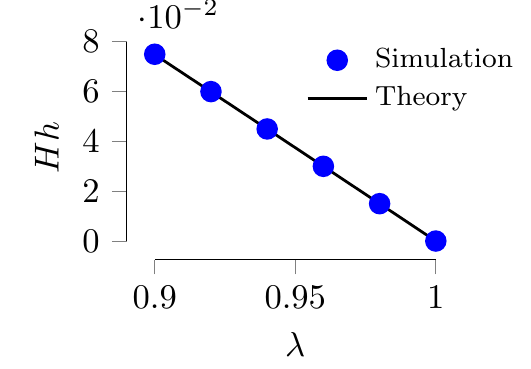}
		\end{tabular}
	\end{center}
	\vspace{-5mm}	
	\caption{$a.$ Numerical simulations of two possible configurations for bilayer prestressed bistable shells. The color map represents the mean curvature of the shell. $b.$, A plot of the mean curvature normalized by the plate's initial thickness as a function of the natural curvature from the applied stimulus; (\protect\tikz \protect\draw[blue,fill=blue] (0,0) circle (.5ex);) numerical results, (\protect\tikz[baseline=-0.5ex] \protect\draw[line width=1pt] (0pt,0pt) -- (5pt,0pt);) equation~\ref{H_kappa_a_b}. $c.$ A plot of the normalized mean curvature as a function of the mechanically applied prestretch $\lambda$; (\protect\tikz \protect\draw[blue,fill=blue] (0,0) circle (.5ex);) numerical results, (\protect\tikz[baseline=-0.5ex] \protect\draw[line width=1pt] (0pt,0pt) -- (5pt,0pt);) equation~\ref{H_final}.}
	\label{fig2}\vspace{-5mm}
\end{figure}

We verified equation~\eqref{H_final} by performing numerical simulations on circular bilayer plates with thicknesses $h \in [1,10]$~mm, radii $R \in [10,100]$~mm, and prestretch $1/\lambda \in [1.01,1.1]$.  The simulations were performed by using the commercial software COMSOL Multiphysics with a neo-Hookean incompressible material model. Plates were made of two layers: the active layer was subjected to a distortion field~$\vett{F}_\textup{$o_1$}=\lambda\vett{e}_1\otimes\vett{e}_1+1\vett{e}_2\otimes\vett{e}_2+\lambda^{-1}\vett{e}_3\otimes\vett{e}_3$, whereas the passive layer was subjected to~$\vett{F}_\textup{$o_2$}=1\vett{e}_1\otimes\vett{e}_1+\lambda\vett{e}_2\otimes\vett{e}_2+\lambda^{-1}\vett{e}_3\otimes\vett{e}_3$, where $(\vett{e}_1,\vett{e}_2,\vett{e}_3)$ is the Cartesian basis. The geometry of the deformed shell was determined by the two natural curvatures $\kappa_{o_1}$ and $\kappa_{o_2}$ in eq.~\ref{H_kappa_a_b}. The dimensionless mean curvature~$Hh$ is measured along the two principal directions of the final prestressed cylindrical shell from the simulation is compared with the analytical results from equations~(\ref{H_kappa_a_b}) and~(\ref{H_final}) in Fig.~\ref{fig2}c \& d.

Having established how to generate a cylindrical shell with a desired mean curvature, we now focus on inducing a snap--through instability of this bistable shell. Snapping between these orientations will occur if the shell's natural curvature is reduced along the cylinder's directrix. From equations~\eqref{kappao} and ~\eqref{kappa_o12}, we know that the natural curvature can be correlated to a change in the stretch of one layer relative to another. Recent work on electrically active polymers has utilized the anisotropic actuation of dielectric elastomers of carbon nanotube electrodes to generate uniaxial stretch in the direction of fiber orientation~\cite{cakmak2015carbon,fang2017enhanced}, so we will examine the response of these bistable shells to the uniaxial stretch along the directrix of one layer. 

\begin{figure} \centering	
	\begin{tabular}{@{}p{0.75\columnwidth}@{\quad}p{width=1\columnwidth}@{}}
		\subfigimg[width=0.35\columnwidth]{$a.$}{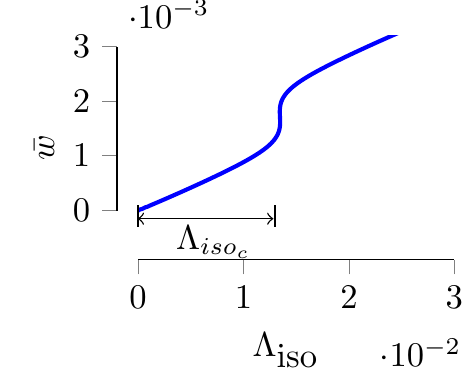}
		\subfigimg[width=0.35\columnwidth]{$b.$}{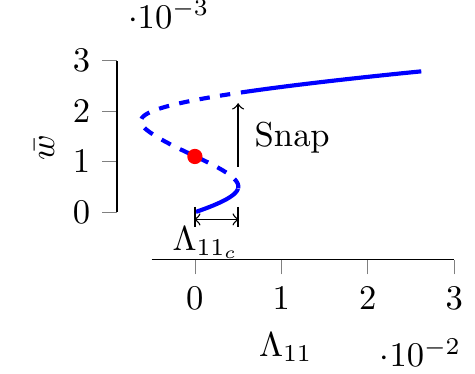}
		\subfigimg[width=0.35\columnwidth]{$c.$}{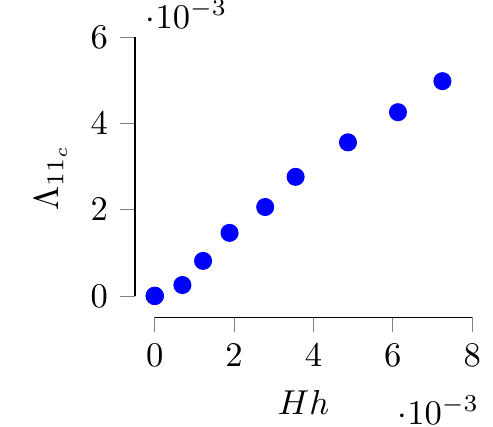}		
		\subfigimg[width=0.35\columnwidth]{$d.$}{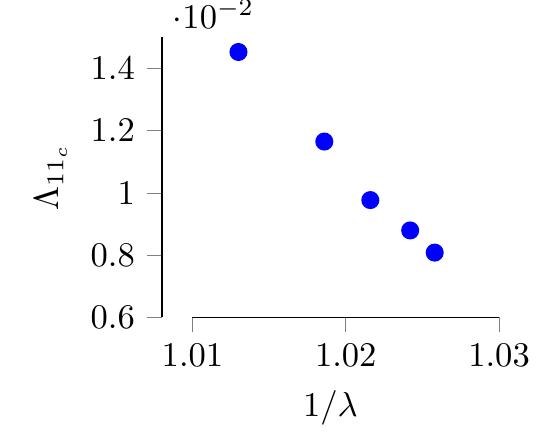}		
	\end{tabular}
	\vspace{-3mm}
	\caption{$a.$ The surface average displacements versus stimuli plot for the snapping of a bilayer cylinder. $b.$ The surface average displacements versus stimuli plot for the snapping of a bilayer prestressed shell. $c.$ The required stimuli to trigger the snapping $\Lambda_{11_c}$ versus $Hh$. $d.$ The required stimuli versus prestretch for shells with fixed $Hh$ and $R$.}
	\label{fig3}
\end{figure}\vspace{0mm}

The bistability of the cylindrical shell was recently considered for stress--free shells subjected to a spherical curvature--inducing stimulus~\cite{pezzulla2017curvature}. In that work, it was assumed that the shell would snap when the total elastic energy in the alternative configuration becomes smaller than the one in the current configuration. Therefore, the critical natural curvature $\kappa_\textup{s}$ needed to induce snapping was obtained by equating the elastic energies in the two configurations. It was shown that $\kappa_\textup{s} \sim 1/(2R_\textup{c})$~\cite{pezzulla2017curvature}. Fig.~\ref{fig3}a. shows the mean displacement of a bilayer cylindrical shell subjected to an isotropic stretch $\Lambda_\textup{iso}$ of one layer, while fig.~\ref{fig3}b. shows the mean displacement of a prestressed cylindrical shell subjected to a uniaxial stretch $\Lambda_{11}$. $\Lambda_{iso_c}$ and $\Lambda_{11_c}$ are the critical stretches to trigger the snapping and are illustrated in fig.~\ref{fig3}. For stress free cylindrical shells, the stretch along the directrix is unable to induce the snapping, therefore we can only use isotropic stretch $\lambda_{iso}$ for them. The dashed curves in the plots represent unstable states that is impossible to achieve during continuously increasing stimuli. By equating the elastic energies before and after snapping, we obtain the inflection point (red dot) of the dashed curves in Fig.~\ref{fig3}. For the snapping of stress--free cylinders, the inflection point of the dashed curves is very close to the critical $\Lambda_\textup{c}$ needed to induce snapping. Therefore, the theoretical value from the energy balance represents a good prediction of the snapping critical states. However, for the snapping of bilayer prestressed shells, the inflection point and actual snapping point are separated by a finite, non-negligible stretch (Fig.~\ref{fig3}). Therefore, to identify the critical stretch of a prestressed shell it is clear that we need to identify an alternative criterion for snap--through.

\begin{figure}
	\begin{center}
		\begin{tabular}{@{}p{0.76\columnwidth}@{\quad}p{width=1\columnwidth}@{}}
			\subfigimg[width=0.75\columnwidth]{$a.$}{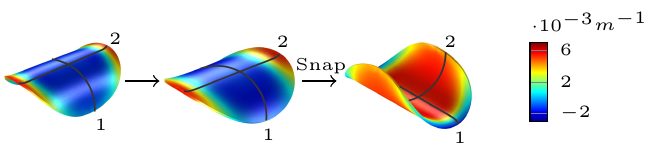}
			\subfigimg[width=0.75\columnwidth]{$b.$}{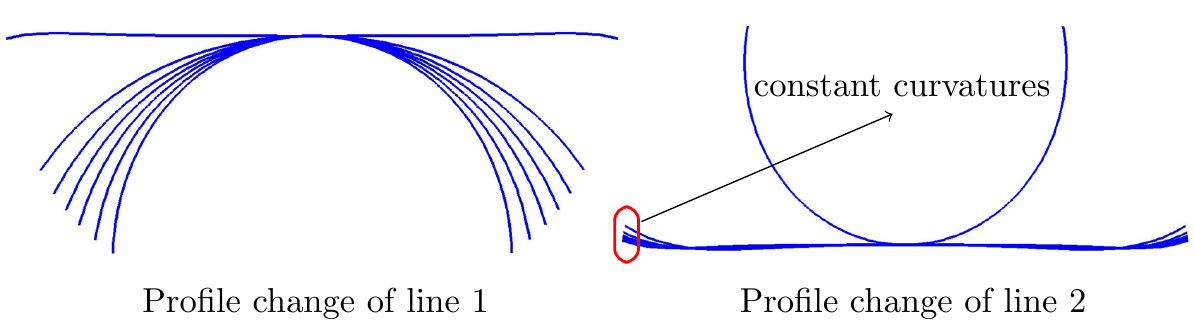}
			\subfigimg[width=0.7\columnwidth]{$c.$}{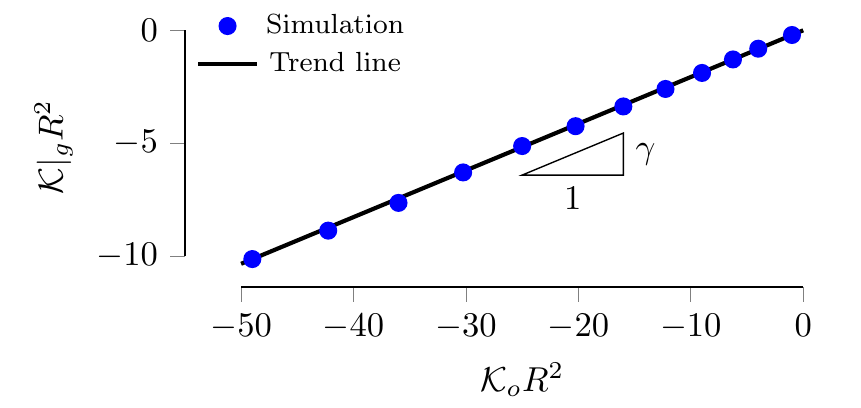}				
		\end{tabular}	
	\end{center} \vspace{-4mm}
	\caption{$a.$The shape change of a prestressed cylindrical shell under external stimuli and after snapping. $b.$ The shape change of the two beams in the mid-surface of the shell during the deformation process under external stimuli. Notice that the curvature at the boundary of line $b$ remain a constant during the deformation process under the external stimuli. $c.$The Gaussian curvature at the boundary is linearly related to the natural Gaussian curvature of the prestressed cylindrical shell. }
	\label{fig4}\vspace{-5mm}
\end{figure}

\begin{figure} \centering
	\begin{tabular}{@{}p{0.75\columnwidth}@{\quad}p{width=1\columnwidth}@{}}
		\subfigimg[width=0.35\columnwidth]{$a.$}{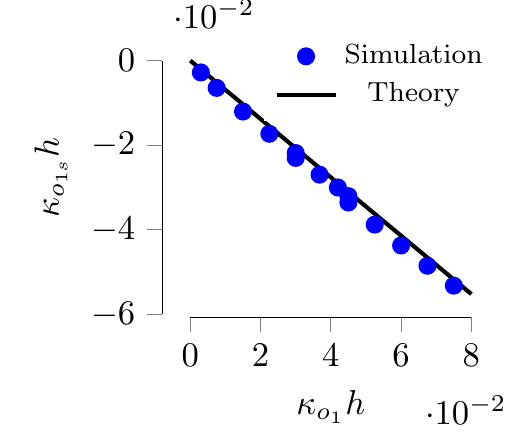}
		\subfigimg[width=0.35\columnwidth]{$b.$}{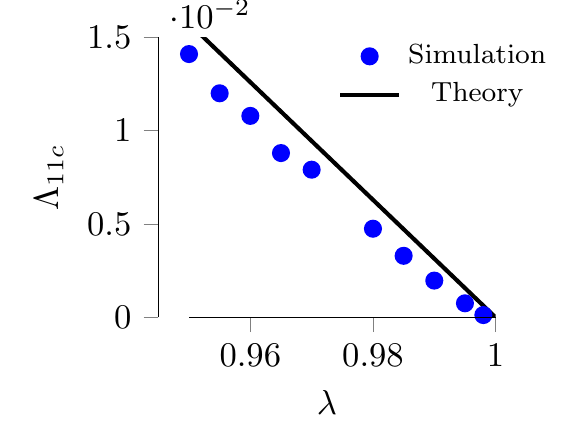}	
	\end{tabular}
	\caption{$a.$ The natural curvature along the directrix when the shell snaps versus the natural curvature along the directrix before the input of external stimuli. $b.$ The external stimuli to trigger the snapping versus prestretch amount.}
	\label{plot5}
\end{figure}

While a prestressed cylinder adopts a mean curvature $\mathcal{H}$ on average, it does not adopt this curvature at every point. As can be seen in Fig.~\ref{fig4}a \& b, while the directrix adopts a curvature of $2\mathcal{H}$, the residual stress causes a portion of the shell boundary along the generatrix to adopt a non--zero curvature. The bulk of the shell adopts an isometry in response to the prestretch, resolving the incompatibility of the applied stimulus and the geometry of the surface. In the boundary, which behaves more like a thin, beam--like region of width $\ell \sim \sqrt{Rh}$, no such incompatibility exists, and so the shell is bending dominated and will try to adopt the natural Gaussian curvature, $\mathcal{K}_{o}=\kappa_{o_1}\kappa_{o_2}$. Therefore, we expect the Gaussian curvature at the end points of the generatrix in the two principal directions of the prestressed cylinder, {\em i.e.} $\mathcal{K}|_\textup{g}=(\kappa_1\kappa_2)|_\textup{g}$, to be proportional to the natural Gaussian curvature, 
\begin{equation} \label{gamma}
	\mathcal{K}|_\textup{g}=\gamma \mathcal{K}_{o} \,,
\end{equation}
in which $\mathcal{K}|_\textup{g}$ is evaluated at a point at the end of the generatrix, and $\gamma$ is a constant that depends on the prestress. Fig.~\ref{plot5}a shows the data points of $\mathcal{K}|_\textup{g}$ and $\mathcal{K}_{o}$ from numerical simulations. The regression line gives~$\gamma\approx0.207$.

As shown in Fig.~\ref{fig4}, we observe that as we gradually apply the stimulus that induces a stretch along the directrix of the prestressed cylindrical shell, the shell will start to unroll while $\kappa_2|_g$ remains at a constant value of $\kappa_2|_\textup{g} \approx -2\mathcal{H}$. We observe numerically that when the shell snaps, the directrix will rotate by $\pi/2$ and will adopt a curvature of $\kappa_2=-2\mathcal{H}$. From the observations that $\mathcal{K}|_\textup{g}=\gamma \mathcal{K}_{o}$ and that at snap--through $\kappa_2=\kappa_2|_\textup{g} = -2\mathcal{H}$, we aim to identify the critical natural curvature needed to induce snap--through. 

To identify the critical snapping curvature, we seek to minimize the strain energy in the boundary layer, while constraining the Gaussian curvature in the boundary layer to remain constant and equal to $\gamma\mathcal{K}_{o}$ with a Lagrange multiplier. We neglect the stretching effect in formulating the strain energy of the shell inside the boundary. The total bending energy in the boundary can be written as
\begin{equation} \label{U_eb}
	\overline{U}|_\textup{g}= \frac{h^2}{3} \int \left[ (1-\nu) \mathrm{tr}(\mathbf{b}-\bar{\mathbf{b}})^2 + \nu \mathrm{tr}^2(\mathbf{b}-\bar{\mathbf{b}}) \right] \id\omega\
\end{equation}
we integrate over a length between $R-\sqrt{Rh} \leq r, \leq R$ $\forall \theta$, where $\mathrm{tr}$ is the trace operator with respect to the natural first fundamental form at the boundary $\bar{\mathbf{a}}$, $\id \omega$ is the relaxed area element that defined as $\sqrt{|\overline{\mathbf{a}}|} \id \eta^1\eta^2$. $|\cdot|$ is the determinant and $\id \eta^\alpha$ are the coordinates along the surface. $\mathbf{b}_\textup{e}$ and $\bar{\mathbf{b}}_\textup{e}$ are the second fundamental form of the shell describing current and natural state respectively. In our problem, $b_{11}=\kappa_1|_\textup{g}$, $b_{22}=\kappa_2|_\textup{g}$, $b_{12}=b_{21}=0$, $\bar{b}_{11}=\kappa_{o_1}|_\textup{g}$, $\bar{b}_{22}=\kappa_{o_2}|_\textup{g}$, $\bar{b}_{12}=\bar{b}_{21}=0$. Based on eq.~(\ref{gamma}), we constrain the Gaussian curvature at the boundary to equal $\gamma \mathcal{K}_{o}$, 
\begin{equation}
	\label{LM}
	\eta\left(\mathcal{K}|_\textup{g}-\gamma \mathcal{K}_{o} \right)=0,
\end{equation}
where $\eta$ is the Lagrange multiplier. The functional to be minimized corresponds to the bending energy augmented by the constraint given by equation~\ref{LM}, and is given by
\begin{equation}
	\label{TPE}
	f(\kappa_1|_\textup{g},\kappa_2|_\textup{g},\eta)= \overline{U}_{b}|_\textup{g}(\kappa_1|_\textup{g},\kappa_2|_\textup{g})-\eta\left(\mathcal{K}|_\textup{g}-\gamma \mathcal{K}_{o} \right)\,.
\end{equation}
Minimization with respect to $\kappa_1|_\textup{g},\kappa_2|_\textup{g}$, and $\eta$ leads to the following equation
\begin{equation} \label{b11b22}
	\nu \kappa_1|_\textup{g}+\kappa_2|_\textup{g}+(1-\nu)\kappa_{o_1}=0\,.
\end{equation}
By solving eq.~(\ref{gamma}) and~(\ref{b11b22}), we get the expressions of $\kappa_1|_\textup{g}$ and $\kappa_2|_\textup{g}$ in terms of $\kappa_{o_1}$, $\nu$ and $\gamma$. Since this boundary curvature is found to be equivalent to $\kappa_{o_2}=\kappa_2|_\textup{g}=-2\mathcal{H}$ at snap-through, the mean curvature for the shell right after snap can be expressed as:
\begin{equation} \label{Hnew1}
	\mathcal{H}_\textup{s}=\frac{1}{4}\left((1-\nu)+\sqrt{(1-\nu)^2+4\nu\gamma}\right) \kappa_{o_2}\,.
\end{equation}
On the other hand, by eq.~(\ref{H_kappa_a_b}), the mean curvature of the shell is also determined by the two natural curvatures along orthogonal directions. Therefore if we use eq.~\ref{H_kappa_a_b} with $\kappa_{o_{1s}}$ to represent the natural curvature along the directrix right before the snapping, then the criteria of the critical snapping state:
\begin{equation} \label{a.snap}
	\kappa_{o_{1s}}=\frac{1}{2\nu}\left[1+\nu-\sqrt{(1-\nu)^2+4\nu\gamma}\right] \kappa_{o_1}\,.
\end{equation} \label{kappa_a.snap}
In the case of incompressible material, i.e. $\nu=0.5$, and with the value of $\lambda$ from eq.~(\ref{gamma}), eq.~(\ref{a.snap}) becomes $\kappa_{o_{1s}}\approx 0.69 \kappa_{o_1}$, which implies that as the prestressed incompressible cylindrical shell unrolls as a result of the uniaxial external stimuli, it will snap to its alternative configuration when the natural curvature decreases to $0.69$ of its original value. Based on this result, we can further establish a relationship between the uniaxial external stimulus $\Lambda_{11_c}$ and the prestretch $\lambda$. In eq.~\ref{a.snap} both $\kappa_{o_1}$ and $\kappa_{o_{1s}}$ are governed by the stretches inside the material, which can be calculated by $\kappa_{o_1}=-\kappa_{o_2}=\mathcal{F}(\lambda)$ that is described in eq.~\ref{kappa_o12}. Therefore by combining eq.~\ref{kappa_o12} and \ref{a.snap} we have the following relationship
\begin{equation} \label{Lambda11c_lambda}
 \mathcal{F}(\lambda+\Lambda_{11_c})=\frac{1}{2\nu}\left[1+\nu-\sqrt{(1-\nu)^2+4\nu\gamma}\right] \mathcal{F}(\lambda).
\end{equation} 
By Taylor expanding the function $\mathcal{F}(\lambda)$ around $1$ up to order $2$, we can solve eq.~\ref{Lambda11c_lambda} and get the relationship between $\Lambda_{11_c}$ and $\lambda$. In the case of $\nu=0.5$ and by using $\gamma=0.207$, the results can be simplified to
\begin{equation}
	\Lambda_{11_c}=0.315(1-\lambda)+\mathcal{O}(\lambda^2),
\end{equation}
The results are verified in fig.~\ref{plot5}c. Eq.~\ref{a.snap} and \ref{Lambda11c_lambda} proposed two ways of looking at the critical snapping conditions for the prestressed shells under uniaxial external stimuli, which are essentially the same.

In this Letter, we focused on a generic stimulus that induces a natural curvature on thin, initially stress--free plates. The natural curvature was correlated to a prestretch $\lambda$ of one layer relative to another, and was used to determine (1.) the shape of resulting residually stressed cylinder, and (2.) the subsequent prestretch required to induce snap--through of the cylinder. Calculation of the shape of the residually stress shell was accomplished using non-Euclidean shell theory, and enabled the precise prediction of a cylinder's mean curvature as a function of the prescribed prestretch. By considering the Gaussian curvature in the boundary layer of the shell, we then identified how the cylinder's geometry and prestress dictates the snap--through criterion. These results will likely inform the design of bistable EAPs, in which prestretching the dielectric elastomers is an important step in fabrication, and the application of a voltage to the EAPs can induce an isotropic or anisotropic stretch in the material depending on the choice of flexible electrode. The ability to use voltage to induce a natural curvature that causes a snap--through instability will help reduce the power required to operate EAPs, and may also reduce the frequency of dielectric breakdown since the voltage will not be required to be applied continuously to maintain a deformation. Typical EAPs are not simple bilayer plates, but will have a more complex multilayer structure. Therefore, additional research is necessary to extend these results to multilayer films with different elastic properties.

\section*{Acknowledgments}
X. Jiang. and D.P. Holmes acknowledge the financial support from NSF through CMMI-1505125.


%

\end{document}